\documentstyle[epsf,multicol,prl,aps]{revtex}
\newcommand{\rmt}{{\rm t}}
\newcommand{\rmd}{{\rm d}}
\newcommand{\rmR}{{\rm r}}
\newcommand{\rmA}{{\rm a}}
\newcommand{\rmRS}{{\rm rs}}
\newcommand{\rmF}{{\rm F}}
\newcommand{\Det}{{\rm Det}}

\newcommand{\rmWZW}{{\rm WZW}}
\newcommand{\calH}{{\cal H}}
\newcommand{\calC}{{\cal C}}
\newcommand{\calT}{{\cal T}}
\newcommand{\calL}{{\cal L}}
\newcommand{\calZ}{{\cal Z}}
\newcommand{\calD}{{\cal D}}
\newcommand{\bmC}{ \mbox{\boldmath $C$} }
\newcommand{\tr}{{\rm tr}}
\newcommand{\bra}[1]{\langle #1|}
\newcommand{\ket}[1]{|#1\rangle}
\newcommand{\tPsi}{{\widetilde\Psi}}
\newcommand{\tpsi}{{\widetilde\psi}}
\newcommand{\tH}{{\widetilde H}}
\newcommand{\bpsi}{{\bar\psi}}
\newcommand{\bchi}{{\bar\chi}}

\newcommand{\slsh}[1]{\!\!\not\! #1}
\begin{document}
\draft
\preprint{}
\title{
Localization-delocalization transition of
disordered $d$-wave superconductors in class $C$I
} 
\author{ 
Takahiro Fukui\cite{Email}
} 
\address{
Department of Mathematical Sciences,
Ibaraki University, Mito 310-8512, Japan
}
\date{January 24, 2000}
\maketitle
\begin{abstract}
A lattice model for disordered $d$-wave superconductors in class
$C$I is reconsidered. Near the band-center, the lattice model can be 
described by Dirac fermions with several species, 
each of which yields WZW term for an effective action of 
the Goldstone mode. The WZW terms cancel out each other 
because of the four-fold symmetry of the model, 
which suggests that the quasiparticle states are localized.
If the lattice model has, however, 
symmetry breaking terms which generate
mass for any species of the Dirac fermions,
remaining WZW term which avoids the cancellation 
can derive the system to a delocalized strong-coupling fixed point.
\end{abstract}


\begin{multicols}{2}

Dirty superconductors have attracted much interest,
since they provide wider universality classes of 
disordered systems \cite{AltZir,Zir,BCSZs}.
In particular, it is quite interesting to ask what the 
universality class of the disordered $d$-wave superconductors
is \cite{NTW,ZHH,SFBN,BCSZd}.
A remarkable property of this unconventional superconductors
is that near the band center, 
quasiparticle states can be described by Dirac fermions \cite{LFSG}.
Such a description enables us, for example, to
relate $d$-wave superconductors with the quantum Hall effect
and to predict a new spin phase called spin quantum Hall fluid 
\cite{SQH}.
They should also produce the well-known effect of
chiral anomaly \cite{PolWie,Wit} to $d$-wave superconductors.
Recently, Senthil {\it et al} \cite{SFBN} have 
studied disordered $d$-wave superconductors with spin rotational 
symmetry and reached the conclusion that all states are localized.
One knows, however, that the WZW term 
due to chiral anomaly plays a crucial role
in two dimensional critical phenomena \cite{Wit}.
Therefore, it is quite important to take the WZW term 
missing in \cite{SFBN} into account,
or to answer the question why it vanishes if 
it does not exist. 

In this paper, we reconsider disorder effects on 
the quasiparticle properties of the $d$-wave superconductors
in the class $C$I 
(those with spin rotational and time-reversal symmetries)
using a replica technique.
It is shown that each Dirac fermion associated with 
four nodes creates the WZW term.
It turns out that they cancel each other and the resultant  
nonlinear sigma model suggests localization 
of the quasiparticles in dirty $d$-wave superconductors,
as was shown by Senthil {\it et al} \cite{SFBN}.
It should be stressed, however, that 
the WZW term is potentially realizable and
the cancellation is accidental:
It is due to the four-fold symmetry of the model.
Therefore, if such symmetry is broken,
the system flows to the strong-coupling fixed point described by
the WZW model. 

Let us begin with a lattice Hamiltonian for singlet
superconductors \cite{NTW,SFBN,BCSZd},
\begin{equation}
H=\sum_{i,j}
\left(
  t_{ij}\sum_{\sigma}c_{i\sigma}^\dagger c_{j\sigma}
 +\Delta_{ij}  c_{i\uparrow}^\dagger   c_{j\downarrow}^\dagger
 +\Delta_{ij}^*c_{j\downarrow}         c_{i\uparrow}
\right).
\label{LatHam}
\end{equation}
We can choose real and symmetric matrices
$t_{ij}=t_{ji}$ and $\Delta_{ij}=\Delta_{ji}$
taking account of the hermiticity  as well as 
the spin-rotational and the time-reversal symmetries.
In the absence of randomness,
we choose the following parameters for 
a pure Hamiltonian $H_0$ with $d$-wave symmetry
\begin{equation}
t_{j,j\pm\hat x}=t_{j,j\pm\hat y}=-t_0, \quad
\Delta_{j,j\pm\hat x}=-\Delta_{j,j\pm\hat y}=\Delta_0,
\end{equation} 
where $\hat x=(1,0)$ and $\hat y=(0,1)$.
Moreover, consider introducing small terms $H_1$  
to break the $C$I symmetry, 
\begin{equation}
\Delta_{j,j+\hat x\pm\hat y}=\Delta_{j+\hat x\pm\hat y,j}=
\pm \frac{i\Delta_1}{4}, \quad
\Delta_{j,j}=i\Delta_1.
\label{LatHam1}
\end{equation}
It will be shown momentarily that this parameter controls the
localization-delocalization transition of the present model.

The pure Hamiltonian $H_0$
has four nodes, where gapless quasi-particle
excitations exist \cite{NTW}. 
Therefore, we can firstly take the 
continuum limit around the nodes of $H_0$ and 
next incorporate the continuum expression of $H_1$,
provided that $H_1$ is small. 
The lattice operators are then described by the 
continuum slowly-varying fields near the band center as\cite{SFBN}, 
\begin{eqnarray}
c_{j\uparrow}/a\sim&&
  i^{ j_x+j_y}\psi_{\uparrow1}^1(x)
 -i^{-j_x-j_y}\psi_{\downarrow2}^1(x) 
\nonumber\\&& 
 +i^{-j_x+j_y}\psi_{\uparrow1}^2(x)
 -i^{ j_x-j_y}\psi_{\downarrow2}^2(x) ,
\nonumber\\
c_{j\downarrow}/a\sim&&
  i^{ j_x+j_y}\psi_{\downarrow1}^{1\dagger}(x)
 +i^{-j_x-j_y}\psi_{\uparrow2}^{1\dagger}(x)  
\nonumber\\&& 
 +i^{-j_x+j_y}\psi_{\downarrow1}^{2\dagger}(x)
 +i^{ j_x-j_y}\psi_{\uparrow2}^{2\dagger}(x) ,
\end{eqnarray}
where $a$ is a lattice constant, $x=aj$, and
two kinds of lower indices of the field $\psi$ are 
referred to as spin, left-right (LR) movers, respectively, 
and upper index as node.
Namely, the field variable $\psi(x)$ lives in the space
$V=\bmC^2\otimes \bmC^2\otimes \bmC^2$.
The pure Hamiltonian in the continuum limit \cite{NTW,SFBN} is then 
$
H_0=
\int d^2x\psi^\dagger(\calH_0+\calH_1)\psi
$
with
\begin{equation}
\calH_0=
\left(
 \begin{array}{ll}
  -\gamma_\mu i\partial_\mu & \\
   &   (x\leftrightarrow y) 
 \end{array}
\right),
\label{ConPurHam}
\end{equation}
where the coordinates have been transformed as 
$x,y\rightarrow\frac{\pm x+ y}{\sqrt{2}}$.
The explicit matrix in Eq. (\ref{ConPurHam}) denotes 
the node space, and 
matrices $\gamma_\mu$ belong to the other space of $V$, 
calculated initially as 
$\gamma_1= v_F 1_2\otimes\sigma_3$ and 
$\gamma_2=v_\Delta 1_2\otimes\sigma_1$,
where $v_F=2\sqrt{2}t_0a$ and $v_\Delta=2\sqrt{2}\Delta_0a$.
It may be more convenient to choose
\begin{equation}
\gamma_1= v_F     1_2\otimes\sigma_2,\quad
\gamma_2=-v_\Delta1_2\otimes\sigma_1,
\end{equation}
via suitable rotation in LR-space of $V$. In this basis,
$\calH_1$ is given by
\begin{equation}
\calH_1=
\left(
 \begin{array}{ll}
   0& \\
   & m1_2\otimes\sigma_3
 \end{array}
\right),
\label{MasTer}
\end{equation}
where $m=\Delta_1$.
Namely, $H_1$ yields asymmetric mass term in the continuum 
Hamiltonian. For the time being we neglect it, but it will be 
shown that vanishing $m$ leads to 
localization whereas finite $m$ drives the system to 
delocalization.

The spin-rotational and time-reversal
symmetries of the lattice model translate, respectively, into the 
continuum model as
\begin{equation}
 \begin{array}{ll}
  \calH=-\calC\calH^\rmt\calC^{-1} , \qquad   &
  \calC=i\sigma_2\otimes\sigma_1\otimes1_2,        \\
  \calH=-\calT\calH\calT^{-1} ,               &
  \calT=1_2\otimes\sigma_3\otimes 1_2,
\end{array}
\label{Sym}
\end{equation}
where t means the transpose.
The total Hamiltonian density
is given by $\calH=\calH_0+\calH_\rmd$,
where $\calH_\rmd$ is disorder 
potential satisfying Eq. (\ref{Sym}).
It is stressed that we take account of 
all kinds of disorder potentials satisfying Eq. (\ref{Sym}). 
This implies that 
the summation with respect to $i,j$
in Eq. (\ref{LatHam}) should be 
over on-site, nearest-neighbor, and diagonal-second-neighbor pairs.
Namely, we can achieve ``maximum information entropy'' 
for the Dirac Hamiltonian with the symmetries (\ref{Sym})
when we introduce 
not only on-site disorder potentials but also 
disordered hopping and off-diagonal pairing for the lattice model
in Eq. (\ref{LatHam}).
Actually Ludwig {\it et al} have derived generic Dirac Hamiltonian
for the integer quantum Hall transition considering lattice model 
in a similar situation \cite{LFSG}. 
It may be straightforward to explicitly calculate disorder 
potentials but tedious to average over them, because
there are no less than twenty independent potentials satisfying
Eq. (\ref{Sym}).
The key point for the ensemble average
is that if the model has ``maximum entropy'', 
we can use the technique developed by Zirnbauer \cite{Zir}.

To study one-quasiparticle properties of the model,
we introduce the Green function
$G_{aa'}(x,x';i\epsilon)=\bra{x,a}(i\epsilon-\calH)^{-1}\ket{x',a'}$,
where index $a$ denotes the set of spin, LR, and node 
species in the space $V$
and $\ket{x,a}=\psi_a^\dagger(x)\ket{0}$. 
Especially, we need
\begin{eqnarray}
&&
G(x)=\sum_a G_{aa}(x,x;i\epsilon),
\nonumber\\
&&
K(x,x')=\sum_{a,a'}G_{aa'}(x,x';i\epsilon)G_{a'a}(x',x;-i\epsilon)
\label{GreFun}
\end{eqnarray}
to compute the DOS and the conductance of the quasiparticle 
transport \cite{NLSM}. 
Some notations are convenient for this purpose.
The introduction of replica for the field $\psi$ 
enables us to express the generating functional 
of these Green functions by path integrals: 
$\psi_{a}\rightarrow\psi_{a\alpha}$ and 
$\psi_{a}^*\rightarrow\psi_{\alpha a}^*$,
where $a$ and $\alpha$ are indices denoting $V$ and 
the replica space $W_\rmR=\bmC^n$, respectively.
The fields $\psi$ and $\psi^*$ have been converted 
into matrix fields, which makes it
simpler to define an order parameter field.
It should be stressed that the fields $\psi$ and $\psi^*$
are completely independent variables.
Lagrangian density is then described symbolically as
$\calL=-\tr_{W_\rmR}\psi^\dagger\left(i\epsilon-\calH\right)\psi$,
where $\tr_{W_\rmR}$ is the trace in the replica space
and the summation over the indices $a$ of the $V$ space is 
implied according to the rule of the matrix product.
It should be noted again that $\psi^\dagger$ is independent of
$\psi$.
Moreover, we introduce an auxiliary space \cite{Zir}
to reflect the symmetries in the $V$ space (\ref{Sym}) 
to an auxiliary field introduced later 
[See Eq. (\ref{SymQ})],
$W_\rmR\rightarrow W=W_\rmR\otimes W_\rmA$ with
$W_\rmA=\bmC^2\otimes\bmC^2$,
which are associated with the 
spin-rotational  and the time-reversal symmetries, respectively.
Fermi fields are now denoted by
$\tPsi_{\alpha i}$ and $\Psi_{i\alpha}$.
One of simpler choices is
\begin{eqnarray}
&&
\Psi=(\psi_+,\psi_-), \quad
\psi_\pm=\calT_\pm\tpsi, 
\nonumber\\
&&
\tPsi=
\left(\begin{array}{c}\bpsi_+\\\bpsi_-\end{array}\right),\quad
\bpsi_\pm=\tpsi^\dagger\calT_\mp, 
\label{DefPsi}
\end{eqnarray}
where 
$\calT_\pm=(1\pm\calT)/2$ 
serves as a projection operator 
( $\calT_++\calT_-=1_2\otimes1_2\otimes1_2$, 
$\calT_\pm^2=\calT_\pm$,
and $\calT_+\calT_-=0$ ) 
into each chiral component of Sp($n$)$\times$Sp($n$) symmetry, 
as we shall see later, and
$\tpsi=\frac{1}{2}\left(\psi_1,-i\psi_2\right)$
with 
$\psi_{1,2}=\psi\pm i\calC^{-1}\psi^*$.
The newly introduced fields $\Psi$ and $\tPsi$
are subject to \cite{Zir},
\begin{equation}
\begin{array}{ll}
 \tPsi=  \gamma \Psi^\rmt \calC^{-1}, \qquad
&\Psi =  \calC \tPsi^\rmt \gamma^{-1},\\
 \tPsi= -\pi   \tPsi      \calT^{-1}, 
&\Psi =  \calT  \Psi      \pi^{-1} .
\end{array}
\label{AuxCon}
\end{equation}
Matrices $\gamma$ and $\pi$ are defined in the $W$ space by
\begin{eqnarray}
&&\gamma=1_n\otimes i\sigma_2\otimes 1_2\equiv
\gamma_0\otimes 1_2 ,
\nonumber\\
&&\pi   =1_n\otimes 1_2      \otimes \sigma_3.
\end{eqnarray}
The identity
$\tr_W(i\epsilon\omega\tPsi\Psi-\tPsi\calH\Psi)
 =\tr_{W_\rmR}\psi^\dagger(i\epsilon-\calH)\psi$,
where 
$\omega=1_n\otimes\sigma_2\otimes\sigma_1$,
leads to the generating functional
$\calZ=\int\!\calD\Psi\calD\tPsi e^{-S}$ with
\begin{equation}
S=-\int\!\!d^2x\tr_W
 \left(
  i\epsilon\omega\tPsi\Psi-\tPsi\calH\Psi+J\tPsi\Psi
 \right).
\label{OrgAct}
\end{equation}

Assume that disorder potential $\calH_\rmd$ obeys the Gaussian 
distribution
$P[\calH_\rmd]=\exp\left(-\frac{1}{2g}\tr_V\calH_\rmd^2\right)$.
Then ensemble average over
disorder is quite simple.
The procedure is as follows\cite{Zir}:
The disorder potentials are integrated out by using the identity
$-\frac{1}{2g}\tr_V\calH_\rmd^2+\tr_V\calH_\rmd\Psi\tPsi
=-\frac{1}{2g}(\tr_V\calH_\rmd-g\Psi\tPsi)^2
+\frac{g}{2}\tr_V(\Psi\tPsi)^2$.
It turns out that the integration over $\calH_\rmd$ is automatic
because $\Psi\tPsi$ satisfy the same symmetries 
as those of $\calH_\rmd$ due to Eq. (\ref{AuxCon}). 
This is actually a merit to consider the disorder potentials
with maximum entropy. If disordered hopping and off-diagonal pairing 
of the lattice model are neglected 
and on-site disorder potentials are merely taken into account, 
some other conditions should be imposed on $\calH_\rmd$ and
therefore on the fields $\Psi$ and $\tPsi$.
Now we have interaction terms of fermions due to ensemble average.
Note the identity 
$\tr_V(\Psi\tPsi)^2=-\tr_W(\tPsi\Psi)^2$.
Then, the four fermi interactions are decoupled via
auxiliary matrix (order parameter)
field defined in the $W$ space.
To be concrete, add the following term into the action,
$\frac{1}{2g}\tr_W (Q+g\tPsi\Psi-\omega)^2$, which is actually
a constant after integration over $Q$.
Then we reach an effective Lagrangian density, 
\begin{eqnarray}
\calL=&&-\tr_W
\left[\frac{1}{2g}\left(Q^2-2i\epsilon\omega Q\right)
+Q\tPsi\Psi-\tPsi\calH_0\Psi\right] .
\label{EffLag}
\end{eqnarray}
Here we have set $J=0$ for simplicity.
Notice that the anti-Hermitian
auxiliary field $Q=-Q^\dagger$ is subject to
\begin{equation}
Q=-\gamma Q^\rmt\gamma^{-1},\quad
Q=-\pi    Q     \pi^{-1} .  
\label{SymQ}
\end{equation}
The solution of these equations is
\begin{equation}
Q=
\left(
 \begin{array}{ll}
   & -M^\dagger\\ M&
 \end{array}
\right)
\label{ExpQ}
\end{equation}
with a condition 
$M= \gamma_0 M^* \gamma_0^{-1}$, where the explicit matrix 
in the above equation denotes the time-reversal space of $W$.

The Lagrangian (\ref{EffLag}) has 
G=Sp($n$)$\times$Sp($n$) symmetry.
To see this, let us consider the transformation   
$Q\rightarrow gQg^{-1}$
which keeps the symmetry relations (\ref{SymQ}).
It turns out that $g$ should satisfy 
$\gamma=g\gamma g^\rmt$ and $\tau g\tau^{-1}=g$
as well as $gg^\dagger=1$, and therefore
$g$ is explicitly given by
\begin{equation}
g=
\left(
 \begin{array}{ll}
  g_+ & \\ & g_-
 \end{array}
\right),
\quad
\label{Expg}
\end{equation}
where $g_\pm\in$ Sp($n$) is a $2n\times2n$  matrix in the replica and 
the spin space of $W$.
So far we have derived the Lagrangian as well as its symmetry group.
To integrate out the fermi fields, it may be convenient to 
use the notations where the fermi fields $\Psi$ and $\tPsi$
are column and row vector, respectively, as usual.
Using Eq. (\ref{SymQ}), fermion part of the Lagrangian (\ref{EffLag})
is rewritten as
$\calL_\rmF
=\tPsi(\calH_0\otimes1-1\otimes Q^\rmt)\Psi
=\tPsi 1 \otimes \gamma
 (\calH_0\otimes1+1\otimes Q)
 1 \otimes \gamma^\dagger\Psi$.
Transform the fields as 
$\tPsi \rightarrow \tPsi 1\otimes\gamma^\dagger$
and 
$\Psi\rightarrow 1\otimes\gamma\Psi$. 
Then the Lagrangian is given by,
in terms of the fields $M$ and $\psi_\pm$,
\begin{eqnarray}
\calL&&=\frac{1}{g}\tr_{W_\rmRS}
\left[
M^\dagger M-\epsilon\gamma_0(M-M^\dagger)
\right]
+\calL_{\rmF1}+\calL_{\rmF2},
\end{eqnarray}
where $W_\rmRS$ is the replica and the spin space of $W$ and
$\calL_{\rmF j}$ describes the Lagrangian of the $j$th-node 
fermion defined by
\begin{equation}
\calL_{\rmF1}=
 \bpsi_+^1 i \slsh{\partial}\psi_+^1
+\bpsi_-^1 i \slsh{\partial}\psi_-^1
-\bpsi_+^1 M^\dagger\psi_-^1 + \bpsi_-^1 M\psi_+^1 ,
\label{LagF1}
\end{equation}
and 
$\calL_{\rmF2}=\calL_{\rmF1}(1\rightarrow2, x\leftrightarrow y)$.
Here and hereafter, the identity matrices such as those in 
$\slsh{\partial}\otimes1$ and  $1\otimes M$ are suppressed.
The transformation laws of $M$ and $\psi_\pm$ fields are
\begin{eqnarray}
&&
M\rightarrow g_-Mg_+^\dagger, \quad
\bpsi_\pm \rightarrow \bpsi_\pm g_\pm^\dagger ,\quad
 \psi_\pm\rightarrow g_\pm \psi_\pm .
\label{TFL}
\end{eqnarray}

We have used a bit complicated basis for $W$ in the definition of
$\Psi$ and $\tPsi$ in Eq. (\ref{DefPsi}),
since $Q$ and $g$ become simpler in this basis.
On the other hand,
it may cause a difficulty in computing the saddle points.
We have known, however, from the $\epsilon$-term in Eqs. 
(\ref{OrgAct}) and (\ref{EffLag}) 
that $\omega$ serves as a ``metric''
in the extended auxiliary space. 
Usually we may choose a basis of the space $W$ with a diagonal metric,
owing to 
which we can assume that $Q$ is also diagonal on the 
saddle points.
In the present case, therefore, it is natural to
assume that $Q$ should have the same structure
as $\omega$, and hence $M_0=v\gamma_0$ with real 
diagonal matrix $v=\mbox{diag}(v_1,\cdots,v_n)$.
Then the variation with respect to $v$ 
after the integration over fermi fields
tells that the saddle points are given by 
$v_\alpha=v_0\sim \Lambda\exp(-\pi v_Fv_\Delta/g)$,
where $\Lambda$ is a ultraviolet cut-off.
This solution gives rise to an exponentially small density of state
at the band-center.
Now it is easy to identify the saddle point manifold as H=Sp($n$):
The chiral transformation $g\equiv(g_+,g_-)$ in 
Eqs. (\ref{Expg}) and (\ref{TFL}) 
is divided into two types. One is 
$g_{\rm v}=(g_1,g_1^*)$ under which $M_0$ is still invariant, 
and the other is 
$g_{\rm a}=(g_2,g_2^\rmt)$ under which $M_0$ is no longer invariant.

Nonlinear sigma model is derived as small fluctuation
around the saddle point manifold H by considering $g_{\rm a}$ type 
local Sp($n$) rotation.
Let us parameterize 
$M=\xi^*H\xi=\tH U$
with $\xi\in$ Sp($n$),
$\tH=\xi^*H\xi^\dagger$, and $U=\xi^2$.
The field $\xi$ describes the massless fluctuation around the 
saddle point manifold H, whereas $H$ describes massive 
longitudinal modes.
In what follows, we take only the leading order for the latter mode,
setting $H=M_0 (=\tH)$.
It should be noted that nonlinear sigma model on G/H has
a global G symmetry as well as a hidden local H symmetry.
Though the field $\xi$ itself is not invariant 
under local H transformation, the composite field $U$ is invariant.

To derive an effective action for the transverse mode, 
let us come back to the Lagrangian (\ref{LagF1}), 
since we should be careful in the integration over fermi fields.
It is practical to firstly integrate out the fermi field of node-1. 
Then, the contribution from node-2 can be obtained by 
replacing $x\leftrightarrow y$. 
To carry out the former integration,
make the transformation
$\bchi_+=  \bpsi_+U^\dagger$ and 
$\chi_+ = U\psi_+$, whereas
$\bchi_-=\bpsi_-1_n\otimes\sigma_2$ and 
$\chi_- =1_n\otimes\sigma_2\psi_-$.
The Lagrangian (\ref{LagF1}) is then converted into
\begin{eqnarray}
\calL_{\rmF1}=
 \bchi_+^1 i \slsh{D}       \chi_+^1
+\bchi_-^1 i \slsh{\partial}\chi_-^1
+iv_0
\left(
\bchi_+^1\chi_-^1 + \bchi_-^1\chi_+^1 
\right),
\end{eqnarray}
where $D_\mu$ is defined by 
$D_\mu=\partial_\mu+L_\mu$ 
with 
$L_\mu=U\partial_\mu U^\dagger$.
It is convenient to scale 
$x=v_Fx'$ and 
$y=v_\Delta y'$ working in the node-1 sector.
Integration over the fermi field of node-1 yields
\begin{eqnarray}
&&
Z_{\rmF1}
=e^{-\frac{1}{2}\Gamma_1(U)}
\nonumber\\&&\times 
\Det^{\frac{1}{2}}_{V_1\otimes W_\rmRS}
1_2\otimes
\left(
\begin{array}{cc}
iv_0
&
i(-i\partial_1-\partial_2)
\\
i( iD_1-D_2)
&
iv_0
\end{array}
\right),
\end{eqnarray}
where the identity matrix $1_2$ belongs to the spin space of $V$,
$V_1$ means the node-1 sector of $V$, 
the derivatives are with respect to the scaled coordinates
$x'$ and $y'$, and
$\Gamma_1(U)$ is the Jacobian for the chiral transformation,
which can be calculated 
by using the Fujikawa method \cite{Fuj}.
We simply present the final answer  
$Z_{\rmF1}\sim e^{-S_1}$, where
the effective action $S_1$ associated with the node-1 fermion
is composed of the principal chiral action of $U$ with a coupling
constant $\lambda=4\pi$ and of the WZW term 
\begin{eqnarray}
\Gamma_{\rmWZW}=
\frac{i}{12\pi}
\int d^3x\epsilon_{\mu\nu\tau}\tr_{W_\rmRS}
\partial_\mu  UU^\dagger
\partial_\nu  UU^\dagger
\partial_\tau UU^\dagger .
\end{eqnarray}
Therefore, 
integration over fermion of node-1 actually yields the WZW
term.

Next let us compute the contribution from the node-2.
The procedure is the rescaling $x'=x/v_F$ and $y'=y/v_\Delta$
and the exchange $x\leftrightarrow y$. 
It should be noted that the WZW term is invariant under 
the rescaling but is odd under the exchange, 
and hence it cancels out.
The total effective action ends up with
\begin{equation}
S=\int d^2x \tr_{W_\rmRS}
\left[
\frac{1}{2\lambda}\partial_\mu U\partial_\mu U^\dagger
-\epsilon
\left(
U+U^\dagger
\right)
\right]
+k\Gamma_{\rmWZW}
\label{EffAct}
\end{equation}
with 
$\frac{1}{\lambda}=\frac{1}{4\pi}\frac{v_F^2+v_\Delta^2}{v_Fv_\Delta}$
and $k=0$.
This is just the action derived by Senthil {\it et al}.
Although the WZW term disappears in the ordinary $d$-wave 
superconductors, we are tempted to expose it,
since the WZW term exists potentially.
This is indeed possible:
If the lattice model includes the symmetry breaking term
(\ref{LatHam1}) and 
therefore the Dirac fermion for the node-2 has a mass
(\ref{MasTer}), we can neglect the node-2 fermion 
in the lower energy than the mass gap. 
In this case, we have the same action (\ref{EffAct}) but with 
$\frac{1}{\lambda}=\frac{1}{4\pi}$ and $k=1$ in the scaled 
coordinates $x'$ and $y'$.

The renormalization group equations of the action (\ref{EffAct})
are calculated at the one-loop order as
\begin{eqnarray}
&&
\frac{d \lambda}{d\ln L}=
-\varepsilon\lambda+\frac{\lambda^2}{4\pi}
\left[
1-
\left(
\frac{k\lambda}{4\pi}
\right)^2
\right],
\nonumber\\
&&
\frac{d \epsilon}{d\ln L}=
\left(
d-\frac{\lambda}{8\pi}
\right)
\epsilon,
\label{RGE}
\end{eqnarray}
where $d=2$, $\varepsilon=d-2$, and
the replica limit $n\rightarrow0$ has been taken.
In the case where $m=0$, we reach the same conclusion as 
Senthil {\it et al}.
Namely, one-quasiparticle states are localized, 
since the spin conductance against weak magnetic fields
is related with $\lambda$ as $\sigma=2/(\pi\lambda)$, which 
is calculated  \cite{SFBN}
via diffusion constant in the diffusion propagator (\ref{GreFun}).
It should be stressed that  
the cancellation of 
the WZW term is due to the four-fold symmetry of the 
$d$-wave Hamiltonian. 
Actually, the latent WZW term can emerge 
via symmetry breaking mass term for the Dirac fermions,
and in that case the coupling constant
$\lambda$ flows to the strong-coupling fixed point value
$\lambda_{\rm c}=4\pi/k$. 
This fixed-point is conformal invariant described by
Sp($n$) WZW model. From the scaling dimension of the energy
in Eq. (\ref{RGE}), it turns out that the density of state 
near this fixed-point obeys the scaling law 
\begin{equation}
\rho(E)=E^{\frac{1}{4k-1}}.
\end{equation}
Therefore, if the pure model has the breaking term 
(\ref{LatHam1}), we suggest that $\rho(E)=E^{\frac{1}{3}}$.

In this paper, we have taken the breaking term of type 
(\ref{LatHam1}) into account. More detailed phase diagram 
will be published elsewhere.
Numerical check of the present conjecture does not seem difficult.
It is quite interesting to expose the hidden WZW term
which exists potentially but conceals itself in the 
four-fold symmetry of the $d$-wave superconductors.

The author would like to thank Y. Hatsugai, Y. Morita,
C. Mudry, and Y. Kato for helpful discussions and comments.
He is deeply indebted to M. R. Zirnbauer and A. Altland 
for valuable discussions in the early stage of this work. 


\end{multicols}

\end{document}